\documentclass[aps,twocolumn,preprintnumbers,superscriptaddress,amsmath,amssymb,footinbib,prl]{revtex4}

\usepackage{epsfig}
\usepackage{graphicx}
\usepackage{bm}
\newcommand{\be}{\begin{equation}}
\newcommand{\ee}{\end{equation}}
\newcommand{\bea}{\begin{eqnarray}}
\newcommand{\eea}{\end{eqnarray}}

\begin{document}
\title{An effective chiral Hadron-Quark Equation of State\\
Part I: Zero baryochemical potential}


\author{J.~Steinheimer}
\affiliation{Institut f\"ur Theoretische Physik, Goethe-Universit\"at, Max-von-Laue-Str.~1,
D-60438 Frankfurt am Main, Germany}
\author{S.~Schramm}
\affiliation{Institut f\"ur Theoretische Physik, Goethe-Universit\"at, Max-von-Laue-Str.~1,
D-60438 Frankfurt am Main, Germany}
\affiliation{Center for Scientific Computing, Max-von-Laue-Str.~1, D-60438 Frankfurt am Main}
\affiliation{Frankfurt Institute for Advanced Studies (FIAS), Ruth-Moufang-Str.~1, D-60438 Frankfurt am Main,
Germany}

\author{H.~St\"ocker}
\affiliation{Institut f\"ur Theoretische Physik, Goethe-Universit\"at, Max-von-Laue-Str.~1,
D-60438 Frankfurt am Main, Germany}
\affiliation{Frankfurt Institute for Advanced Studies (FIAS), Ruth-Moufang-Str.~1, D-60438 Frankfurt am Main,
Germany}
\affiliation{GSI Helmholtzzentrum f\"ur Schwerionenforschung GmbH, Planckstr.~1, D-64291 Darmstadt, Germany}

\begin{abstract}
We construct an effective model for the QCD equation of state, taking
into account chiral symmetry restoration as well as the deconfinement phase transition.
The correct asymptotic degrees of freedom at the high and low temperature limits are
included (quarks $\leftrightarrow$ hadrons). The model shows a rapid crossover for both order parameters, as is expected from lattice calculations. We then investigate the thermodynamic properties of
the model at $\mu_B=0$. All thermodynamic quantities are in qualitative agreement with lattice data, while apparent quantitative differences can be attributed to hadronic contributions and excluded volume corrections.
\end{abstract}

\maketitle

\section{Introduction}
The recent experimental results at the Relativistic Heavy Ion Collider (RHIC), suggesting the creation of a nearly perfect fluid
\cite{Adams:2005dq,Back:2004je,Arsene:2004fa,Adcox:2004mh},
 have fueled interest in the study of bulk properties of strongly interacting matter (QCD).
 Heavy ion experiments at different beam energies try to map out the QCD phase diagram, especially
 the region where one expects a phase transition from a confined gas of hadrons to a deconfined state of
 quarks and gluons (QGP)
 \cite{Gyulassy:2004zy,Fodor:2001pe,Fodor:2007vv,Karsch:2004wd,Stephanov:1998dy,Gazdzicki:1998vd,Stephanov:1999zu,Bravina:1999dh,Bravina:2000dk,Gazdzicki:2004ef,Arsene:2006vf,Theis:1984qc}. To relate any experimental observables to the properties of the matter produced in heavy ion collisions, a profound understanding of the thermodynamics of QCD has to be obtained and integrated in model simulations of these collisions.\\
One can define two different phase transitions, the first being the chiral phase transition associated with chiral symmetry restoration in the vanishing quark mass limit, where the chiral condensate serves as a well defined order parameter. In the limit of heavy quarks, a deconfinement phase transition with the Polyakov loop as order parameter, is assumed. Physical quarks however have intermediate masses and one would expect that at least the deconfinement order parameter is not so well defined anymore. There could even be some mixing of the two order parameters, accounting for some lattice QCD observation that deconfinement and chiral restoration occur at the same temperature (at least at $\mu_B=0$) \cite{Fukugita:1986rr,Karsch:1994hm,Aoki:1998wg,Karsch:2000kv,Allton:2002zi}.\\  
Such lattice calculations at finite temperature are an important tool for the investigation of the QCD phase diagram.
For the thermodynamics of the pure gauge theory high accuracy data is available \cite{Boyd:1996bx}, and the equation of state (EoS) of strongly interacting matter at vanishing chemical potential is reasonably well understood \cite{Karsch:2000ps,Aoki:2006br}.
Here lattice predicts a rapid crossover for the deconfining and chiral phase transitions.\\
At finite baryo-chemical potential, lattice calculations suffer from the so called sign problem. There are several
different approaches to obtain results at
finite $\mu_B$ \cite{Fodor:2002km,Allton:2002zi,Allton:2003vx,Allton:2005gk,de Forcrand:2003hx,Laermann:2003cv,D'Elia:2004at,D'Elia:2002gd}, but yet no clear picture, especially about the existence and location of a possible critical end point, has emerged.\\
Recent considerations based on connecting the large $N_c$ limit with real-word QCD
draw an even more exotic picture of the phase diagram,
where the critical temperatures of the deconfinement and chiral phase transitions disconnect and depart
in the region of high net baryon densities \cite{Hidaka:2008yy}.\\
In our approach we combine, in a single model, a well-established flavor-SU(3) hadronic model with a quark-gluon description of the highly
excited matter. This allows us to study the chiral-symmetry and confinement-deconfinement phase structure of the strongly interacting matter at high temperatures and densities. In addition we obtain an equation of state of hadronic and quark matter
that is applicable over a wide range of thermodynamical conditions and that can therefore be used in heavy-ion simulations with
very different beam energies.

\section{Model description}
In our approach we derive the EoS of hot and dense nuclear matter using a single model for the hadronic and quark phase.
The model includes the correct asymptotic degrees of freedom, namely a free gas of quarks and gluons at
infinite temperature, and a gas of hadrons having the correct vacuum properties at vanishing temperature. The model also predicts the structure of finite nuclei, nuclear and neutron matter properties and a first order liquid-vapor phase transition.
The two phase transitions that are expected from QCD, the chiral and deconfinement transitions, are
also included in a consistent manner.\\
In the following we will show how we describe the different phases of QCD and how we combine them in a single model.\\

We describe the hadronic part of the EoS, using a flavor-SU(3) model which
is an extension of a non-linear representation of a sigma-omega model including
the pseudo-scalar and vector octets of mesons and the baryonic octet and decuplet
(for a detailed discussion see \cite{Papazoglou:1998vr,Papazoglou:1997uw,Dexheimer:2008ax}).

The Lagrangian density of the model in mean field approximations reads:
\begin{eqnarray}
&L = L_{kin}+L_{int}+L_{meson},&
\end{eqnarray}
where besides the kinetic energy term for hadrons, the terms:

\begin{eqnarray}
&L_{int}=-\sum_i \bar{\psi_i}[\gamma_0(g_{i\omega}\omega+g_{i\phi}\phi+m_i^*]\psi_i,&
\end{eqnarray}
\begin{eqnarray}
&L_{meson}=-\frac{1}{2}(m_\omega^2 \omega^2+m_\phi^2\phi^2)\nonumber&\\
&-g_4\left(\omega^4+\frac{\phi^4}{4}+3\omega^2\phi^2+\frac{4\omega^3\phi}{\sqrt{2}}+\frac{2\omega\phi^3}{\sqrt{2}}\right)\nonumber&\\
&+\frac{1}{2}k_0(\sigma^2+\zeta^2)-k_1(\sigma^2+\zeta^2)^2&\nonumber\\
&-k_2\left(\frac{\sigma^4}{2}+\zeta^4\right)-k_3\sigma^2\zeta&\nonumber\\
&+ m_\pi^2 f_\pi\sigma -k_4\ \frac{\chi^4}{\chi_0^4} \ln{\frac{\sigma^2\zeta}{\sigma_0^2\zeta_0}} &\nonumber\\
&+ \chi^4-\chi_0^4 + \ln\frac{\chi^4}{\chi_0^4}~.&
\label{formel1}
\end{eqnarray}
represent the interactions between baryons
and vector and scalar mesons, the self-interactions of
scalar and vector mesons, and an explicitly chiral symmetry breaking term.
The index $i$ denotes the baryon octet and decuplet. Here, the mesonic condensates (determined in
mean-field approximation) included are
the vector-isoscalars $\omega$ and $\phi$, and
the scalar-isoscalars $\sigma$ and $\zeta$ (strange quark-antiquark state). Assuming isospin symmetric matter, we can neglect the $\rho$-meson contribution in Eq. \ref{formel1}.

The last four terms of (\ref{formel1}) were introduced to model the QCD trace anomaly \cite{Papazoglou:1997uw}, where the dilaton field $\chi$ can be identified with the gluon condensate.

The effective masses of the baryons (of the octet)
are generated by the scalar mesons except for an explicit
mass term ($\delta m_N=120$ MeV):
\begin{eqnarray}
&m_{b}^*=g_{b\sigma}\sigma+g_{b\zeta}\zeta+\delta m_b,&
\end{eqnarray}
while, for simplicity and in order to reduce the number of free parameters, the masses of the decuplet baryons are kept at their vacuum expectation values.
With the increase of temperature/density, the $\sigma$ field (non-strange chiral condensate) decreases
its value, causing the effective masses of the particles to decrease towards chiral symmetry restoration.
The coupling constants for the baryons \cite{Dexheimer:2009hi} are chosen
to reproduce the vacuum masses of the baryons, nuclear saturation properties and
asymmetry energy as well as the $\Lambda$-hyperon potentials. The vacuum expectation values of the scalar
mesons are constrained by reproducing the pion and kaon decay constants.\\

The extension of the hadronic SU(3) model to quark degrees of freedom
is constructed in analogy to the PNJL model \cite{Fukushima:2003fw,Ratti:2005jh}.
The sigma model uses the Polyakov loop $\Phi$ as the order parameter for
deconfinement. $\Phi$ is defined via $\Phi=\frac13$Tr$[\exp{(i\int d\tau A_4)}]$, where $A_4=iA_0$ is the temporal component
of the SU(3) gauge field. One should note that one must distinguish $\Phi$, and its conjugate $\Phi^{*}$
at finite baryon densities \cite{Fukushima:2006uv,Allton:2002zi,Dumitru:2005ng}, as they couple differently to the quarks, respective antiquarks.\\
In our approach the effective masses of the quarks
are generated by the scalar mesons except for a small explicit
mass term ($\delta m_q=5$ MeV and $\delta m_s=105$ MeV for the strange quark):
\begin{eqnarray}
&m_{q}^*=g_{q\sigma}\sigma+\delta m_q,&\nonumber\\
&m_{s}^*=g_{s\zeta}\zeta+\delta m_s,&
\end{eqnarray}
with values of $g_{q\sigma}=g_{s\zeta}= 4.0$. At present
we do not consider any possible quark-vectormeson couplings.

A coupling of the quarks to the Polyakov loop is introduced in the thermal energy of the quarks. Their thermal contribution to the grand canonical potential $\Omega$, can then be written as:
\begin{equation}
	\Omega_{q}=-T \sum_{i\in Q}{\frac{\gamma_i}{(2 \pi)^3}\int{d^3k \ln\left(1+\Phi \exp{\frac{E_i^*-\mu_i}{T}}\right)}}
\end{equation}
and
\begin{equation}
	\Omega_{\overline{q}}=-T \sum_{i\in Q}{\frac{\gamma_i}{(2 \pi)^3}\int{d^3k \ln\left(1+\Phi^* \exp{\frac{E_i^*+\mu_i}{T}}\right)}}
\end{equation}

The sums run over all quark flavors, where $\gamma_i$ is the corresponding degeneracy factor, $E_i^*$ the energy and $\mu_i$ the chemical potential of the quark.\\
All thermodynamical quantities, energy density $e$, entropy density $s$ as well as the
densities of the different particle species $\rho_i$, can be derived from the grand canonical potential. In our model it has the form:
\begin{equation}
	\frac{\Omega}{V}=-L_{int}-L_{meson}+\frac{\Omega_{th}}{V}-U
\end{equation}
Here $\Omega_{th}$ includes the heat bath of hadronic and quark quasi particles. The effective potential $U(\Phi,\Phi^*,T)$ which controls the dynamics of the Polyakov-loop will be discussed in the following.
In our approach we adopt the ansatz proposed in \cite{Ratti:2005jh}:
\begin{eqnarray}
	U&=&-\frac12 a(T)\Phi\Phi^*\nonumber\\
	&+&b(T)ln[1-6\Phi\Phi^*+4(\Phi^3\Phi^{*3})-3(\Phi\Phi^*)^2]
\end{eqnarray}
 with $a(T)=a_0 T^4+a_1 T_0 T^3+a_2 T_0^2 T^2$, $b(T)=b_3 T_0^3 T$.\\

This choice of effective potential satisfies the $Z(3)$ center symmetry of the pure gauge Lagrangian. In the confined phase, $U$ has an absolute minimum at $\Phi=0$, while above the critical Temperature $T_0$ (for pure gauge $T_0 = 270$ MeV) its minimum is shifted to finite values of $\Phi$. The logarithmic term appears from the Haar measure of the group integration with respect to the SU(3) Polyakov loop matrix. The parameters $a_0, a_1, a_2$ and $b_3$ are fixed, as in \cite{Ratti:2005jh}, by demanding a first order phase transition in the pure gauge sector at $T_0=270$ MeV, and that the Stefan-Boltzmann limit is reached for $T \rightarrow \infty$.
Note that $T_0$ remains a free parameter to adjust the actual critical temperature,
of both phase transitions, when both, quarks and hadrons, couple to the scalar fields.\\

As has been mentioned above, the Lagrangian of the chiral model contains dilaton terms to model the scale anomaly. These terms constrain the chiral condensate, if the dilaton is frozen at its ground state value $\chi_0$. On the other hand, as deconfinement is realized, the expectation value of the chiral condensate should vanish at some point. On account of this we will couple the Polyakov loop to the dilaton in the following way:
\begin{equation}
	\chi^2=\chi_0^2 \ (1-(\phi\phi^*)^2)
\end{equation}
 
Hence, when the value of the Polyakov loop approaches unity, the dilaton field will slowly vanish and allow the chiral condensate to also approach zero.
\\

Until now all hadrons are still present in the deconfined and chirally restored phase.
Since we expect them to disappear, at least at some point above $T_c$, we have to include
a mechanism that effectively suppresses the hadronic degrees of freedom, when deconfinement is achieved.\\
In previous calculations baryons were suppressed by introducing a large baryon mass shift for
non-vanishing $\Phi$ \cite{Dexheimer:2009hi}.

In the following the suppression mechanism will be provided by excluded volume effects.
It is well known that hadrons are no point-like particles, but have a finite volume.
Including effects of finite-volume particles, in a thermodynamic model for hadronic matter, was proposed some time ago \cite{Hagedorn:1980kb,Baacke:1976jv,Gorenstein:1981fa,Hagedorn:1982qh}. We will use an ansatz similar to that used in \cite{Rischke:1991ke,Cleymans:1992jz}, but modify it to also treat the point like quark degrees of freedom consistently.\\

\begin{figure}[t]
\centering
\includegraphics[width=0.5 \textwidth]{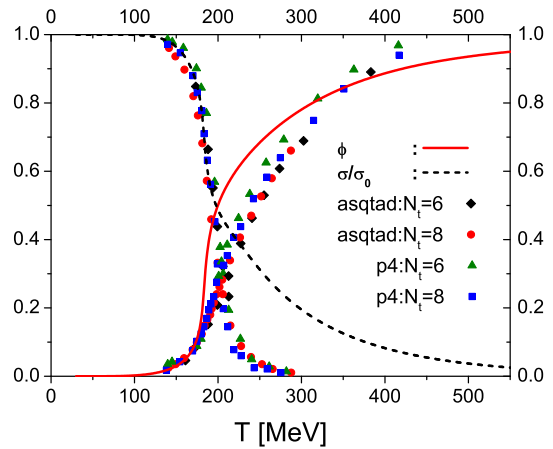}
\caption{\label{mu0sig}
(color online) The normalized order parameters for the chiral (red, solid line), and deconfinement (black, dashed line) phase transition as a function of $T$ at $\mu_B = 0$. The symbols denote lattice data from \cite{Bazavov:2009zn}, using different lattice actions (asqdat and p4) and lattice spacings $N_{\tau}$.}
\end{figure}

If one introduces a particle of radius $r$ into a gas of the same particles, then the volume excluded is not just the simple spherical volume, but one-half times the volume of a sphere with radius $2r$:
\begin{equation}
	v=\frac12 \cdot \frac43 \pi (2r)^3
\end{equation}
 It is easy to understand that if all other particles also have a radius $r$ then the excluded volume is much bigger than just the volume of a single particle.\\
We expect the volume of a meson to be smaller than of a baryon we have to introduce the quantity $v_{i}$ which is the volume excluded of a particle of species $i$. Since we only distinguish between hadronic baryons, mesons and quarks. Consequently  $v_{i}$ can only assume three values:
\begin{eqnarray}
 v_{Quark}&=&0 \nonumber \\
 v_{Baryon}&=&v \nonumber \\
 v_{Meson}&=&v/a \nonumber \\
\end{eqnarray}
where $a$ is a number larger than one. In our calculations we assumed it to be $a=8$, which would mean that the radius $r$ of a meson is half that of a baryon.
Note that we neglect any possible Lorentz contraction effects on the excluded volumes as introduced in \cite{Bugaev:2000wz,Bugaev:2008zz}.

The modified chemical potential $\widetilde{\mu}_i$ which is connected to the real chemical potential $\mu_i$, of the $i$-th particle species, is obtained by the following relation:
\begin{equation}
	\widetilde{\mu}_i=\mu_i-v_{i} \ P
\end{equation}
where $P$ is the sum over all partial pressures. All thermodynamic quantities can then be calculated with respect to the temperature $T$ and the new chemical potentials $\widetilde{\mu}_i$. To be thermodynamically consistent, all densities ($\widetilde{e_i}$, $\widetilde{\rho_i}$ and $\widetilde{s_i}$) have to be multiplied by a volume correction factor $f$, which is the ratio of the total volume $V$ and the reduced volume $V'$, not being occupied,:
\begin{equation}
	f=\frac{V'}{V}=(1+\sum_{i}v_{i}\rho_i)^{-1}
\end{equation}

The actual densities then are:
\begin{eqnarray}
e&=&\sum_i f \ \widetilde{e_i} \\
\rho_i&=&f \ \widetilde{\rho_i} \\
s&=&\sum_i f \ \widetilde{s_i}
\end{eqnarray}

Note that in this configuration the chemical potentials of the hadrons are decreased by the quarks, but not vice versa. In other words as the quarks start appearing they effectively suppress the hadrons by changing their chemical potential, while the quarks are only affected through the volume correction factor $f$.\\
Our implementation of finite-volume corrections
as outlined above is a simple approach with as few parameters as possible
and can be improved upon in various ways. For one, hadrons differ in size.
The size of a hadron could even be density or temperature dependent \cite{Kapusta:1982qd}. In addition, the excluded-volume parameter
of a particle does also depend on the density of the system (at dense packing a particle excludes
effectively less volume). However, one should regard the variables $v$ and $a$ as effective parameters
for capturing the qualitative effect of an excluded volume correction, which suppresses the hadrons in the quark phase.\\

\section{Results at vanishing net baryon density}

\begin{figure}[t]
\centering
\includegraphics[width=0.5 \textwidth]{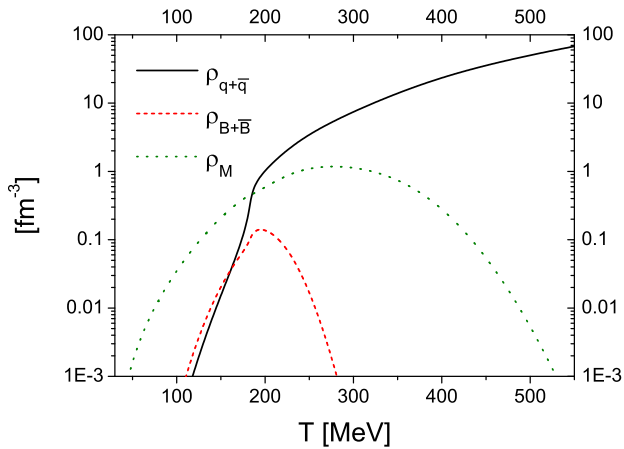}
\caption{\label{mu0dens}
(color online) Total particle number densities for the different particle species as a function of $T$ at $\mu_B = 0$. The black line shows the total number of quarks+antiquarks per volume while the green (dotted) line refers to the total meson density and the red (dashed) line to the number density of hadronic baryons+antibaryons.}
\end{figure}

In this paper we will concentrate on the properties of the model at vanishing chemical potential. Here different lattice calculations suggest a crossover from the hadronic to the quark phase. Different lattice groups obtain different results for the phase transition temperature ranging from $T_c= 160$ MeV to $200$ MeV \cite{Aoki:2009sc,Detar:2007as}.
For all following results we set $T_0$, the free parameter of the Polyakov-potential, to $T_0=235$ MeV and the excluded volume parameter $v= 1 {\rm fm}^3$. This leads to a critical temperature of $T_c \approx 183$ MeV ($T_c$ is defined as the temperature with the largest change in the order parameters as a function of the temperature).\\
The lattice data referred to in the following sections are taken from the HotQCD collaboration \cite{Bazavov:2009zn}. Here different actions (p4, asqtad) and lattice spacings ($N_{\tau}= 6,8$) were compared. Note that the transition region extracted from the lattice data lies between $185$ and $195$ MeV.\\
Fig. \ref{mu0sig} shows the temperature dependence for the order parameters of both,
the deconfinement ($\Phi$), and chiral ($\sigma$) phase transition, extracted from our model and compared to lattice data. Both order parameters change smoothly with temperature. The critical temperature is found to be equal for both
phase transitions . Note that the value of the chiral condensate $\sigma$ only very slowly approaches zero. This originates from the dilaton contribution to the scalar potential in this model, which includes a repulsive term for small values of $\sigma$.

\begin{figure}[t]
\centering
\includegraphics[width=0.5 \textwidth]{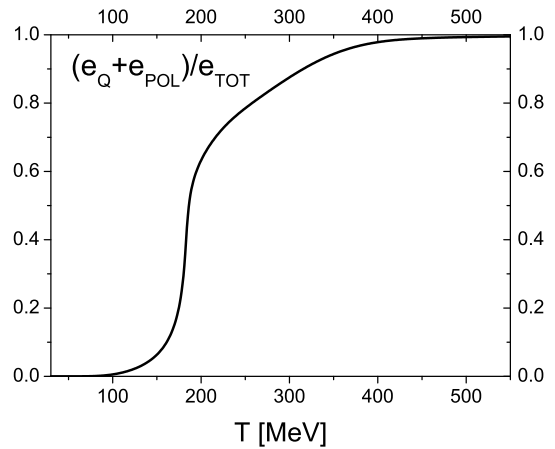}
\caption{\label{mu0mix}
The fraction of the total energy density that can be assigned to the quark-gluon phase ($e_{QGP}$ contains the energy of the quarks and the Polyakov potential) as a function of $T$ at $\mu_B = 0$.}
\end{figure}

Fig. \ref{mu0dens} shows the total densities of quarks plus antiquarks (black solid line), mesons (green dotted line) and baryons plus antibaryons (dashed red line). Below the critical temperature hadrons are the dominant degree of freedom. When the quark number increases around $T_c$, they begin to suppress the hadrons. It is remarkable that the hadrons are still present, and not negligible, up to about $2.0 \ T_c$ \cite{Stoecker:1980uk}. Especially the mesons contribute strongly to all thermodynamic quantities, since they are quite less suppressed than the baryons ($v_M < v_B$). Above $2 \ T_c$ the hadrons are effectively squeezed out of the system by the presence of the quarks.

To emphasize this change in degrees of freedom, Fig. \ref{mu0mix} shows the fraction of the total energy density which stems from the quarks and gluons (more precisely the Polyakov potential). As expected for a crossover both degrees of freedom (hadrons and quarks) are present in the temperature range from $0.75-2 \ T_c$. Around $T_c$ the fraction of the energy density, due to quarks and gluons increases rapidly. It converges to unity at around 2 times $T_c$.

\begin{figure}[t]
\centering
\includegraphics[width=0.5 \textwidth]{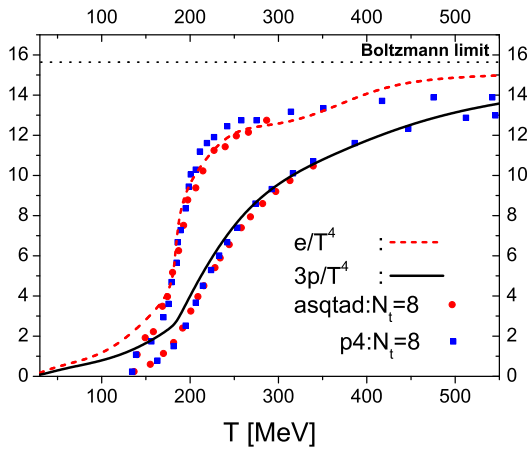}
\caption{\label{mu0et4}
(color online) Three times the pressure (red dashed line) and energy density (solid line) over $T^4$ as a function of $T$ at $\mu_B = 0$. The green dotted line indicates the Boltzmann limit for an ideal gas of three massless quarks and gluons. The symbols denote lattice data from \cite{Bazavov:2009zn}, using different lattice actions (asqdat and p4) and lattice spacings $N_{\tau}$.}
\end{figure}

Let us now take a closer look at different thermodynamic quantities. Fig  \ref{mu0et4} displays the energy density (black curve) and three times the pressure (red dashed curve), both over $T^4$ compared to lattice data \cite{Bazavov:2009zn}. In the limit of infinite temperature, both quantities should converge to the Stefan-Boltzmann limit of an ideal gas of quarks and gluons. This limit is indicated as a green dashed line. The strong increase in energy density around $T_c$ reflects the rapid change of the relevant degrees of freedom. At three times the critical temperature the energy density is slowly converging to the Stefan-Boltzmann limit, while the pressure is converging even slower as it was also observed in PNJL calculations \cite{Roessner:2006xn}. At temperatures below $T_c$ our calculation gives larger values for the pressure and the energy density, which is not surprising as the correct description of the hadron contribution is problematic in the case of lattice calculations \cite{Bazavov:2009zn}.\\

\begin{figure}[t]
\centering
\includegraphics[width=0.5 \textwidth]{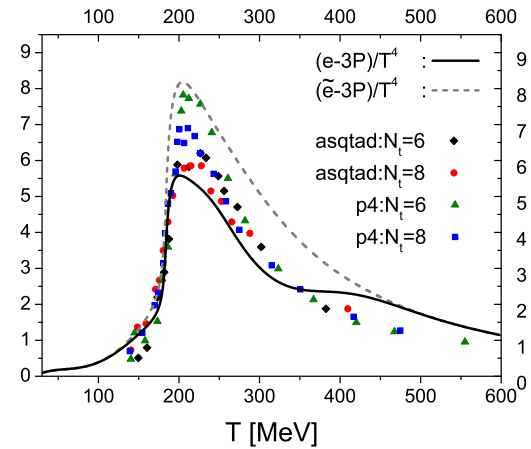}
\caption{\label{mu0e3p}
(color online) Energy density minus three times the Pressure over $T^4$ as a function of $T$ at $\mu_B = 0$. Also referred to as the interaction measure.  The dashed line depicts the interaction measure using the uncorrected energy density $\widetilde{e}$. The symbols denote lattice data from \cite{Bazavov:2009zn}, using different lattice actions (asqdat and p4) and lattice spacings $N_{\tau}$.}
\end{figure}

\begin{figure}[b]
\centering
\includegraphics[width=0.5 \textwidth]{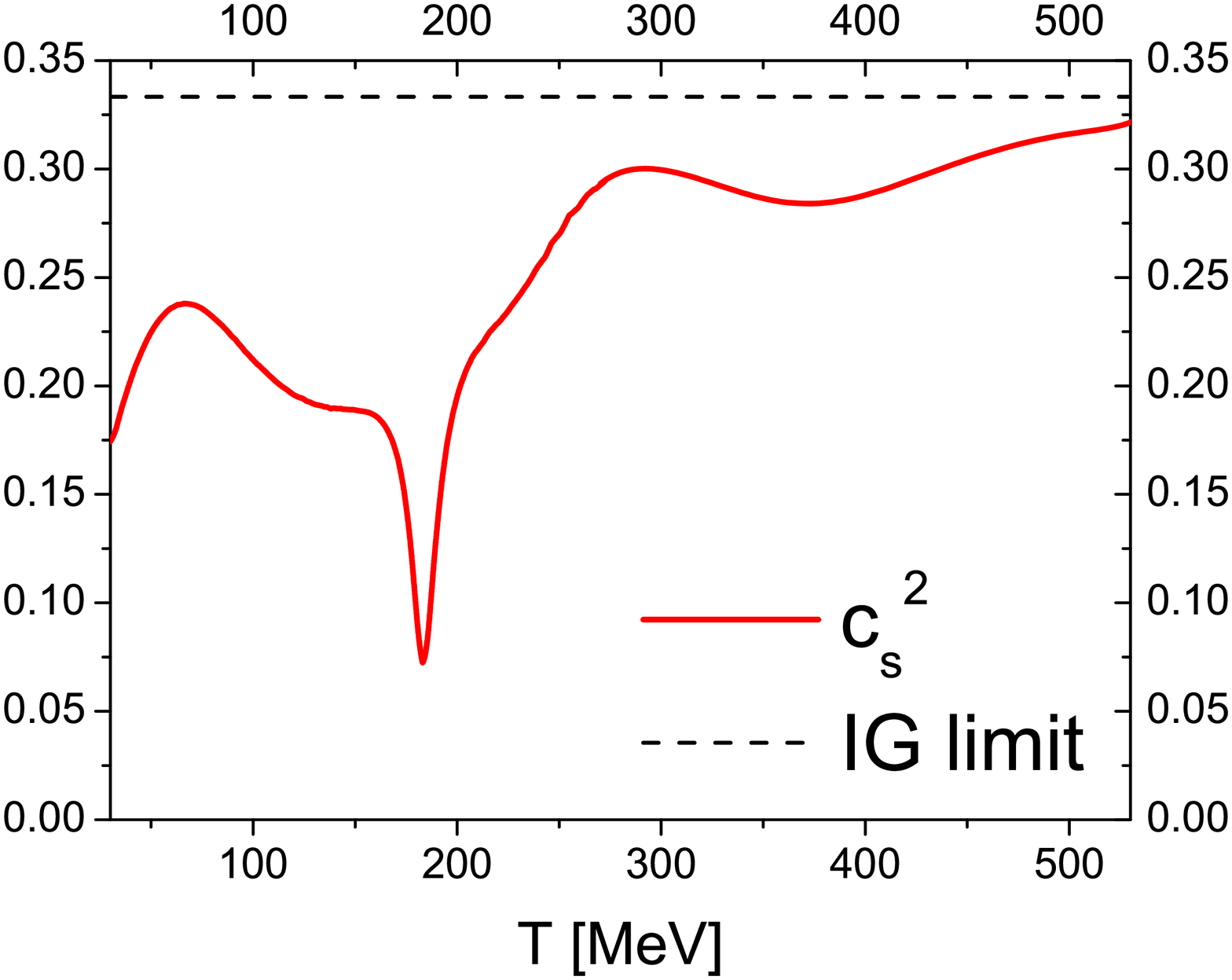}
\caption{\label{mu0cs2}
(color online) The speed of sound squared, as a function of $T/T_c$ at $\mu_B = 0$. $1/3$ is the ideal gas limit.}
\end{figure}

Around $1.5 \ T_c$ one can observe a slight 'dip' in the energy density. This 'dip' is connected to the correction factor $f$ of the excluded volume corrections. As has been shown above, the hadronic contribution to the densities disappears only at two time $T_c$ and therefore they still exclude some portion of the volume for the quarks. The 'dip' therefore indicates the disappearance of volume correction factors for the quark phase.

In the high temperature limit, where only the quarks (and gluons) remain in the system the energy density and pressure both slightly exceed the data from lattice calculations.

Figure \ref{mu0e3p} displays the difference of the energy density and three times the pressure over $T^4$ (black solid line). This quantity is also referred to as the 'interaction measure' in lattice calculations. In the
Stefan-Boltzmann limit it is $0$, while it shows a peak slightly above $T_c$. The height of the peak in our model is comparable to the lower bound from lattice studies \cite{Bazavov:2009zn}, while its value at large $T$ is a little bit above that from lattice calculations, because chiral restoration is not fully achieved in our model. To point out the effect of the volume correction factor on this quantity, Figure \ref{mu0e3p} also displays the uncorrected interaction measure $(\widetilde{e}-3p)/T^4$ as a grey dashed line.

An important property of a hot and dense nuclear medium is the speed of sound ($c_s$):
\begin{equation}
c_s^2 = \left. \frac{d p}{d e}\right|_{\mu=0}
\end{equation}
It is not only closely related to expansion dynamics but also controls the way perturbations (sound- and shock-waves) travel
through the fireball \cite{Stoecker:1980vf}. Fig \ref{mu0cs2} shows the speed of sound squared as a function of temperature. As the temperature increases towards $T_c$ one can clearly observe a softening of the EoS due to the crossover. At very high temperature the speed of sound again converges toward its ideal gas limit of $c_s^2 \rightarrow 1/3$. The dip above $T_c$ is again related to the excluded volume corrections. Note that even though the change of degrees of freedom from hadrons to quarks proceeds as a crossover, there is still a substantial softening (i.e. $c_s^2$ goes down to $0.07$)!

\section{Extrapolation to finite baryo chemical potentials}

\begin{figure}[t]
\centering
\includegraphics[width=0.5 \textwidth]{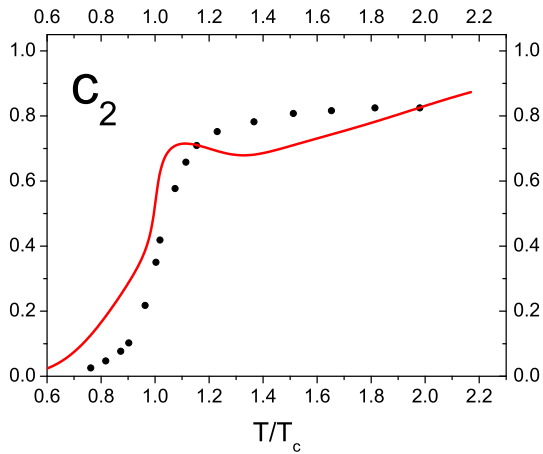}
\caption{\label{mu0susz}
(color online) The Taylor coefficient $c_2$ (red, solid line) as a function of $T/T_c$ at $\mu_B = 0$. The symbols denote lattice data from \cite{Allton:2005gk}.}
\end{figure}

\begin{figure}[t]
\centering
\includegraphics[width=0.5 \textwidth]{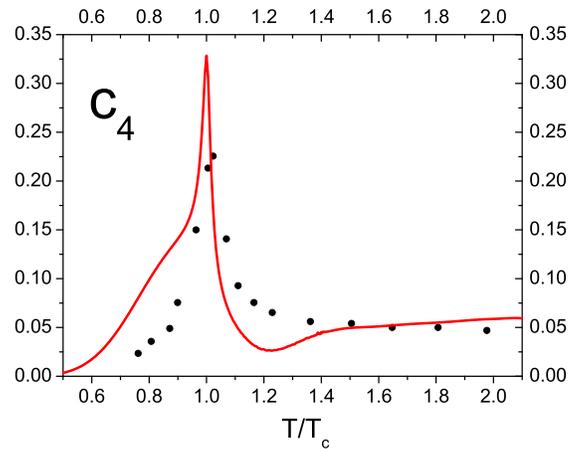}
\caption{\label{mu0suszc4}
(color online) The Taylor coefficient $c_4$ (red, solid line) as a function of $T/T_c$ at $\mu_B = 0$. The symbols denote lattice data from \cite{Allton:2005gk}.}
\end{figure}

Lattice results at finite chemical potentials ($\mu_B=3 \mu_q \neq 0$) are often obtained as Taylor expansion of the thermodynamic quantities in the
parameter $\mu/T$ around zero chemical potential and $T = T_c$. We perform the same expansion in the present model to compare with lattice results. In the Taylor expansion of the pressure $p=-\Omega/V$ with respect to $\frac{\mu_q}{T}$, the coefficients $c_n$ follow from:
\begin{eqnarray}
\frac{p(T,\mu_q)}{T^4}&=& \sum_{n=0}^{\infty}c_n(T)\left(\frac{\mu_q}{T}\right)^n \\
c_n(T)&=&\left. \frac{1}{n!} \frac{\partial^n(p(T,\mu_q)/T^4)}{\partial(\mu_q/T)^n}\right|_{\mu_q=0}
\end{eqnarray}

Because the derivatives of the pressure with respect to $\mu_q/T$ are difficult to be obtained directly from a mean field approach (see \cite{Ghosh:2006qh}), we calculate the pressure as function of $T$ and $\mu_q$ explicitly and then extract the derivatives around $\mu_q = 0$.\\

The results for the second moment are shown in Fig \ref{mu0susz} and compared to lattice calculations taken from \cite{Allton:2005gk}. The second order coefficient $c_2$, exhibits an enhancement below $T_c$, due to the hadronic contributions and a minimum just above $T_c$. This is not observed in lattice calculations, because the origin of this dip is the presence of the additional hadronic degrees of freedom and the excluded volume effects in the hadronic part of the model. Thus it is not surprising to see a discrepancy to the lattice data.\\
The fourth order coefficient $c_4$ is shown in Fig \ref{mu0suszc4}. As expected, a peak is observed around $T_c$. Again the absolute height of the peak exceeds that from lattice calculations. This, again, stems from the fact that our model includes many hadronic degrees of freedom around $T_c$. The minimum above $T_c$ is also visible in the forth order coefficient.\\
Note that the quark number susceptibility is also closely related to the Taylor coefficients and therefore shows a clear maximum at $T_c$.\\

As our model does not sustain the difficulties of lattice calculations, when going to finite net baryon densities, it is straight forward to extend our investigations to finite chemical potentials ($\mu_B=  3 \mu_q \neq 0$). These calculations at finite densities are in progress and will be subject of future publications.\\ 

\section{Conclusion}

We present a novel approach for modeling an EoS that respects the symmetries underlying QCD,
and includes the correct asymptotic degrees of freedom, i.e. quarks and gluons at high temperature and hadrons in the low-temperature limit. We achieve this by including quarks degrees of freedom and the thermal
contribution of the Polyakov loop in a hadronic chiral sigma-omega model. The hadrons are suppressed at high
densities by excluded volume corrections. Nevertheless, we observe a substantial hadronic contribution to the EoS up to temperatures of 2 times $T_c$.\\
We can show that the properties of the EoS are in qualitative agreement with lattice data at $\mu_B = 0$. Various quantities, like the pressure and energy density, are in good agreement with lattice data. Deviations from lattice results ,
for example the Taylor coefficients of the expansion in $\mu/T$, can be explained by the hadronic contributions and volume corrections.
In spite of a continuous phase transition, we obtain a considerably smaller value for the speed of sound around $T_c$ ($c_s^2 \approx 0.072$) when compared to lattice calculations \cite{Bazavov:2009zn}.

\section*{Acknowledgments}
This work was supported by BMBF, HGS-hire and the Hessian LOEWE initiative through the Helmholtz International center for FAIR (HIC for FAIR). The authors thank D. Rischke, C. Greiner, M. Bleicher, J. Noronha, V. Dexheimer and P. Koch-Steinheimer for fruitful discussions. The computational resources were provided by the Frankfurt Center for Scientific Computing (CSC).

\end{document}